\documentclass[conference]{IEEEtran}
\IEEEoverridecommandlockouts

\usepackage{cite}
\usepackage{amsmath,amssymb,amsfonts}
\usepackage{graphicx}
\usepackage{booktabs}
\usepackage{xcolor}
\usepackage{listings}
\usepackage{textcomp}
\usepackage{url}
\usepackage{tikz}
\usetikzlibrary{arrows.meta,fit,backgrounds,shadows}
\usepackage{pgfplots}
\pgfplotsset{compat=1.17}
\usepackage[hidelinks]{hyperref}

\definecolor{mitRed}{HTML}{750014}
\definecolor{mitBrightRed}{HTML}{FF1423}
\definecolor{mitSilverGray}{HTML}{8B959E}
\definecolor{mitBlack}{HTML}{000000}
\definecolor{mitWhite}{HTML}{FFFFFF}
\colorlet{bandFront}{mitWhite!94!mitSilverGray}\colorlet{accFront}{mitRed}
\colorlet{bandIR}{mitWhite!88!mitSilverGray}   \colorlet{accIR}{mitSilverGray!85!mitBlack}
\colorlet{bandOpt}{mitWhite!86!mitBrightRed}   \colorlet{accOpt}{mitBrightRed!80!mitRed}
\colorlet{bandRt}{mitWhite!82!mitSilverGray}   \colorlet{accRt}{mitRed!70!mitBlack}
\colorlet{bandBk}{mitWhite!78!mitSilverGray}   \colorlet{accBk}{mitBlack!70!mitSilverGray}
\colorlet{bandPf}{mitWhite!74!mitBrightRed}    \colorlet{accPf}{mitBrightRed}

\definecolor{codegray}{rgb}{0.5,0.5,0.5}
\definecolor{codekw}{rgb}{0.10,0.10,0.55}
\definecolor{codestr}{rgb}{0.58,0.0,0.0}
\lstdefinestyle{rlxrust}{
  basicstyle=\ttfamily\scriptsize,
  keywordstyle=\color{codekw}\bfseries,
  commentstyle=\color{codegray}\itshape,
  stringstyle=\color{codestr},
  showstringspaces=false,
  columns=fullflexible,
  keepspaces=true,
  breaklines=true,
  numbers=none,
  morekeywords={let,mut,fn,use,vec,version,features,assert_eq},
  comment=[l]{//},
  morestring=[b]",
}
\begin{document}

\title{RLX: A Unified Multi-Backend Tensor Compiler\\and Distributed Runtime in Rust}

\author{%
\IEEEauthorblockN{Eugene Hauptmann}
\IEEEauthorblockA{\textit{Massachusetts Institute of Technology} \\
Cambridge, MA, USA \\
eugenehp@mit.edu}
\and
\IEEEauthorblockN{Nataliya Kosmyna}
\IEEEauthorblockA{\textit{Massachusetts Institute of Technology} \\
Cambridge, MA, USA \\
nkosmyna@mit.edu}
}

\maketitle

\begin{abstract}
Production machine learning (ML) stacks often split graph compilation and kernel 
execution across different layers and languages, making backend behavior, 
deployment guarantees, and performance fallbacks hard to reason about end-to-end. 
RLX addresses this gap with a single Rust codebase that combines compiler and 
runtime roles around one primitive-level, three-level intermediate representation (IR), 
plus a transparent dispatch contract that resolves each operator to native,
common-IR, or rewritten lowering and fails compilation when legalization is not
possible. The same IR targets fourteen runtime devices (\texttt{cpu},
\texttt{metal}, \texttt{mlx}, \texttt{ane}, \texttt{cuda}, \texttt{rocm},
\texttt{oneapi}, \texttt{tpu}, \texttt{hexagon}, \texttt{gpu},
\texttt{vulkan}, \texttt{opengl}, \texttt{directx}, \texttt{webgpu}) and
two specialty codegen paths (Cortex-M INT8 and FPGA), ingests safetensors,
GGUF, ONNX, and rten formats, supports F16/BF16/F64/C64 and quantized INT4/INT8
flows with AMP/PTQ/QAT, and scales via tensor-/pipeline-parallel collectives
over TCP and RDMA transports. Beyond neural workloads, RLX also extends to
scientific/physics-style domains through sparse and dense linear algebra
extensions (e.g., CSR LU/CG/matvec and LAPACK-backed factorizations) and
3D Gaussian splatting operators. We evaluate RLX against PyTorch, TensorFlow, JAX,
candle, burn, tch, rten, MLX, CoreML, IREE, Glow, TensorRT, and tinygrad under
identical input generation and p50 measurement methodology on one host. On
\texttt{all-MiniLM-L6-v2}, RLX-Metal is fastest at every batch (e.g.,
$16.6$\,ms at batch~32 vs. PyTorch-MPS $26.7$\,ms). In the MNIST training
table, RLX also has the top-throughput entry (graph-fused MLP:
946{,}487\,img/s), above NumPy${+}$BLAS (787{,}349\,img/s), while retaining
100\% top-1 parity on reference checks (e.g., Qwen3).
\end{abstract}


\section{Introduction}
The progress of machine learning has been coupled to the ease with which
developers can turn high-level model descriptions into efficient code
on diverse hardware. Two strategies dominate. \emph{Graph
compilers} such as XLA~\cite{xla}, TVM~\cite{tvm}, and
MLIR~\cite{mlir} capture a whole-program representation and apply
fusion, layout, and scheduling transformations before emitting code.
\emph{Kernel runtimes} such as cuDNN~\cite{cudnn}, Apple's Metal
Performance Shaders, and MLX~\cite{mlx} instead provide a curated set of
hand-optimized primitives that an eager front end dispatches at run time.
The first strategy maximizes cross-operator optimization; the second
minimizes the distance between a call and a vendor-tuned kernel.

In practice the two are combined together with adapter layers,
foreign-function boundaries, and a host language, typically Python, that
is separate from the systems language in which the kernels are written.
Python's global interpreter lock, per-object reference counting, and
dynamic dispatch impose a latency cost that adds up at small batch
sizes and limits threading. It further complicates deployment
to limited or non-traditional targets, such as microcontrollers, web browsers,
custom silicon, and makes it hard to verify end-to-end if the code
will actually run.

We introduce RLX a single, memory-safe system language which can
host \emph{both} a graph compiler and a kernel runtime, with one IR
spanning the full range of targets from a server GPU down to an INT8
microcontroller. RLX is implemented as a Rust workspace of layered
crates. Its IR is a small, composable set of primitives: 
structured control flow (\texttt{Op::Scan}), dense linear solves
(\texttt{Op::DenseSolve}), quantization (\texttt{Op::FakeQuantize},
\texttt{Op::DequantMatMul}), masked attention with an explicit
\texttt{MaskKind}, and a first-class custom-op extension surface. The
optimizer performs fusion, automatic-mixed-precision policy, reverse- and
forward-mode automatic differentiation, batching (\texttt{vmap}),
broadcast legalization, and post-training-quantization insertion. The
runtime performs backend dispatch, compile caching, and cost-based device
selection across heterogeneous hardware.

This paper contributes a unified IR, optimizer, and runtime in one Rust
workspace (Section~\ref{sec:design}); a transparent dispatch that fails
and does not silently degrade (Section~\ref{sec:dispatch}) the graph ops and performance; portability across
fourteen runtime and two codegen targets (Section~\ref{sec:backends}); 
extensibility, including distributed TP/PP (Section~\ref{sec:cases},
Section~\ref{sec:distributed}); and a cross-framework evaluation with runnable
benchmark scripts (Section~\ref{sec:eval}).

\section{Background and Motivation}
\label{sec:background}
\subsection{Requirements}
We designed RLX using four requirements:
(i)~\emph{heterogeneous targets} – one system reaching multicore CPUs,
Apple unified-memory GPUs, discrete NVIDIA/AMD GPUs, TPUs, neural
accelerators, and edge microcontrollers and custom logic, without a rewrite
per device and responsive to per-deployment objectives (throughput,
precision, IO latency, and power);
(ii)~\emph{one language without host/kernel split} – the IR, optimizer,
runtime, and kernels in one memory-safe language, so a model and its
lowering are inspectable from a single tree;
(iii)~\emph{transparency over silent fallback} – the system reports, per
operator, natively, or if it is lowered, or rewritten path and
\emph{fails} when an operator cannot be legalized for the chosen device;
and (iv)~\emph{extensibility from primitives} – quantization schemes,
custom operators, and downstream domains (sparse and dense linear algebra,
Gaussian splatting) expressed against a stable extension surface.

\subsection{Motivation behind choosing Rust}
Rust gives memory and thread safety without garbage collection, a build
system (Cargo) that scales to a large workspace, and zero-cost FFI to vendor
libraries (cuBLAS, cuDNN, MPSGraph, MLX, libtpu) - so RLX keeps the IR,
optimizer, and backend kernels in one tree while still calling hand-tuned
vendor code where it wins. The Rust frameworks Burn~\cite{burn} and
Candle~\cite{candle} show community interest in this direction; RLX differs
in coupling a primitive-level compiler IR to a transparent dispatch contract and
in targeting MCU and FPGA from the same IR.

\subsection{Related work}
RLX draws its representation from JAX~\cite{jax} and XLA~\cite{xla}
(composable array primitives, differentiation and batching as
transformations), its whole-program ethos from TVM~\cite{tvm},
MLIR~\cite{mlir}, IREE~\cite{iree}, and Glow~\cite{glow}, and its
ergonomics from PyTorch~\cite{pytorch}. The Rust frameworks
candle~\cite{candle}, burn~\cite{burn}, tch~\cite{tch} and the engines
rten~\cite{rten} and ONNX~Runtime~\cite{ort} are the closest neighbors;
retargetable stacks (Mojo~\cite{mojo}, TensorRT~\cite{tensorrt},
tinygrad~\cite{tinygrad}, Luminal~\cite{luminal}, ZML~\cite{zml}) share
the compile-an-IR philosophy. MLIR is infrastructure for \emph{building}
compilers – no runtime, kernels, or autodiff of its own (IREE supplies
those) – whereas RLX is a complete system with all
three layers. Section~\ref{sec:comparison} makes the axis-by-axis comparison.

\section{System Design}
\label{sec:design}
\begin{figure*}[t]
\centering
\resizebox{0.78\textwidth}{!}{%
\begin{tikzpicture}[
  font=\footnotesize,
  comp/.style={draw=black!45, rounded corners=2.5pt, fill=white,
               align=center, inner sep=2.5pt,
               drop shadow={shadow xshift=0.4pt, shadow yshift=-0.5pt,
                            opacity=0.35, fill=black!55}},
  cA/.style={comp, text width=22mm, minimum height=8mm},
  cIR/.style={comp, text width=36mm, minimum height=9mm},
  cOpt/.style={comp, text width=40mm, minimum height=12mm},
  cRt/.style={comp, text width=37mm, minimum height=12mm},
  cBk/.style={comp, text width=11mm, minimum height=7.2mm, inner sep=1.4pt, font=\scriptsize},
  cPf/.style={comp, text width=22mm, minimum height=8mm},
  band/.style={rounded corners=4pt, line width=0.7pt},
  badge/.style={circle, text=white, font=\scriptsize\bfseries,
                inner sep=0pt, minimum size=4.6mm},
  blab/.style={font=\footnotesize\bfseries, align=center, text width=16mm},
  flow/.style={-{Stealth[length=2.6mm,width=2.6mm]}, line width=1.4pt,
               draw=black!50},
  spine/.style={-{Stealth[length=1.6mm,width=1.6mm]}, line width=0.6pt,
               draw=black!40},
  tag/.style={font=\scriptsize\itshape, text=black!55},
]
\draw[band, fill=bandFront, draw=accFront] (-0.15, 9.88) rectangle (19.15,11.12);
\draw[band, fill=bandIR,    draw=accIR]    (-0.15, 7.95) rectangle (19.15, 9.35);
\draw[band, fill=bandOpt,   draw=accOpt]   (-0.15, 5.95) rectangle (19.15, 7.55);
\draw[band, fill=bandRt,    draw=accRt]    (-0.15, 3.90) rectangle (19.15, 5.60);
\draw[band, fill=bandBk,    draw=accBk]    (-0.15, 2.03) rectangle (19.15, 3.67);
\draw[band, fill=bandPf,    draw=accPf]    (-0.15, 0.60) rectangle (19.15, 1.70);
\node[badge, fill=accFront] at (0.5,10.5){1}; \node[blab, text=accFront] at (1.7,10.5){Front ends \& formats};
\node[badge, fill=accIR]    at (0.5, 8.65){2}; \node[blab, text=accIR]    at (1.7, 8.65){IR \texttt{rlx-ir}};
\node[badge, fill=accOpt]   at (0.5, 6.75){3}; \node[blab, text=accOpt]   at (1.7, 6.75){Optimizer \texttt{rlx-opt}};
\node[badge, fill=accRt]    at (0.5, 4.75){4}; \node[blab, text=accRt]    at (1.7, 4.75){Runtime \texttt{rlx-runtime}};
\node[badge, fill=accBk]    at (0.5, 2.85){5}; \node[blab, text=accBk]    at (1.7, 2.85){Backends};
\node[badge, fill=accPf]    at (0.5, 1.15){6}; \node[blab, text=accPf]    at (1.7, 1.15){Platforms};
\draw[spine] (9.5,9.88) -- (9.5,9.35);  \node[tag] at (9.78,9.62){HIR};
\draw[spine] (9.5,7.95) -- (9.5,7.55);  \node[tag] at (9.78,7.75){MIR};
\draw[spine] (9.5,5.95) -- (9.5,5.60);  \node[tag] at (9.78,5.78){MIR};
\draw[spine] (9.5,3.90) -- (9.5,3.67);  \node[tag] at (9.78,3.79){LIR};
\draw[spine] (9.5,2.03) -- (9.5,1.70);
\node[cA] at (3.80,10.5){Tensor DSL\\(lazy trace)};
\node[cA] at (6.62,10.5){Graph\\builder};
\node[cA] at (9.44,10.5){pyrlx\\(Python)};
\node[cA] at (12.26,10.5){safetensors};
\node[cA] at (15.08,10.5){GGUF\\v1--v3};
\node[cA] at (17.90,10.5){ONNX\\import};
\draw[dashed,accFront!70] (10.85,9.95) -- (10.85,11.05);
\node[tag] at (6.6,9.62){author}; \node[tag] at (15.0,9.62){ingest weights};
\node[cIR] at (6.20,8.65){\textbf{HIR}: block ops\\(Linear, SwiGLU)};
\node[cIR] at (11.00,8.65){\textbf{MIR}: fused\\tensor DAG};
\node[cIR] at (15.80,8.65){\textbf{LIR}: DAG +\\buffer plan};
\draw[flow, line width=1pt] (8.05,8.65) -- (9.15,8.65);
\draw[flow, line width=1pt] (12.85,8.65) -- (13.95,8.65);
\node[tag] at (8.60,8.92){lower}; \node[tag] at (13.40,8.92){plan};
\node[cOpt] at (5.10,6.75){\textbf{rlx-fusion}\\region fusion $\cdot$ unfuse};
\node[cOpt] at (10.85,6.75){\textbf{rlx-autodiff}\\grad $\cdot$ jvp $\cdot$ hvp\\$n$th-order $\cdot$ vmap};
\node[cOpt] at (16.60,6.75){\textbf{rlx-compile}\\legalize $\cdot$ mem-plan\\AMP $\cdot$ PTQ $\cdot$ QAT};
\node[cRt] at (4.90,4.75){Session\\+ compile cache};
\node[cRt] at (8.87,4.75){Transparent dispatch\\native $\!\to\!$ common-IR\\$\to$ rewrite $\to$ fail};
\node[cRt] at (12.83,4.75){Device router\\(cost-based)};
\node[cRt] at (16.80,4.75){Distributed collectives\\TP / PP\\TCP / RDMA};
\node[cBk] at (3.00,2.85){CPU};
\node[cBk] at (4.30,2.85){Metal};
\node[cBk] at (5.60,2.85){MLX};
\node[cBk] at (6.90,2.85){CUDA};
\node[cBk] at (8.20,2.85){ROCm};
\node[cBk] at (9.50,2.85){oneAPI};
\node[cBk] at (10.80,2.85){TPU};
\node[cBk] at (12.10,2.85){Hexagon};
\node[cBk] at (13.40,2.85){GPU APIs};
\node[cBk] at (14.70,2.85){ANE\\(CoreML)};
\node[cBk] at (16.00,2.85){Cortex-M};
\node[cBk] at (17.30,2.85){FPGA};
\draw[dashed,accBk!70] (15.35,2.10) -- (15.35,3.30);
\node[tag] at (8.85,3.46){runtime targets (Device + legalization contract)};
\node[tag] at (16.65,3.46){specialty codegen};
\node[cPf] at (3.85,1.15){macOS / iOS};
\node[cPf] at (6.65,1.15){Linux / Windows};
\node[cPf] at (9.45,1.15){Android};
\node[cPf] at (12.25,1.15){Web\\(WebGPU/WASM)};
\node[cPf] at (15.05,1.15){Embedded\\(MCU $\cdot$ FPGA)};
\node[cPf] at (17.85,1.15){ASIC\\(via rlx-eda)};
\end{tikzpicture}%
}
\caption{RLX architecture. Front ends and weight-format loaders
(safetensors, GGUF, ONNX) trace into one primitive-level IR defined
in \texttt{rlx-ir} over a single operator set (113 kinds). IR has
three levels, named on the vertical edges by the representation crossing
them: \textbf{HIR} (high-level block ops, e.g.\ \texttt{Linear},
\texttt{SwiGLU}) lowers to \textbf{MIR} (the fused primitive tensor DAG the
optimizer consumes), which is buffer-planned into \textbf{LIR} (the input
each backend lowers). The same IR reaches fourteen runtime targets (with
GPU variants exposed as \texttt{gpu}/\texttt{vulkan}/\texttt{opengl}/\texttt{directx}/\texttt{webgpu})
plus two specialty codegen targets and every major platform.}
\label{fig:arch}
\end{figure*}
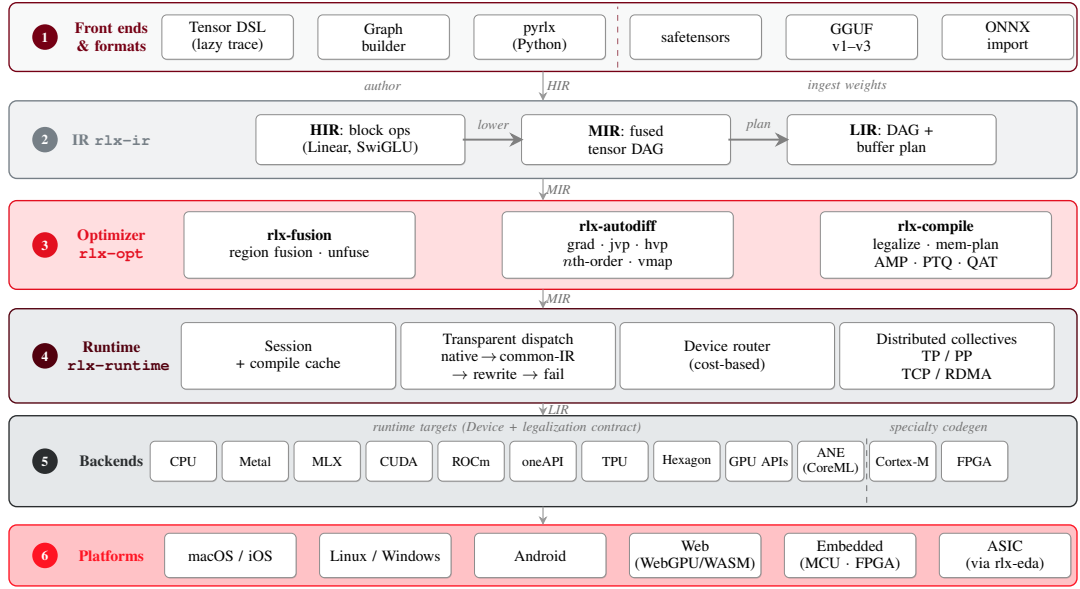

RLX is organized as three layers in a IR crate, shown in
Figure~\ref{fig:arch}: authoring front ends and weight-format ingestion
trace into a single IR, which an optimizer and runtime lower, dispatch, and
distribute across a wide set of hardware backends and deployment platforms.

\subsection{Intermediate representation}
The crate \texttt{rlx-ir} defines the tensor IR: an \texttt{Op}
enum, \texttt{Shape} and \texttt{DType} types with a promotion table, a
graph builder, shape inference, and a verifier. The IR is deliberately
primitive-level. It contains roughly 113+ operators, including
structured control flow (\texttt{Scan}), reductions, masked attention
parameterized by \texttt{MaskKind}~$\in\{$\texttt{None},
\texttt{Causal}, \texttt{SlidingWindow}$(w)$, \texttt{Custom}$\}$, the
recurrent family (\texttt{Gru}, \texttt{Rnn}, \texttt{Mamba2}), Fourier
transforms (\texttt{Op::Fft}, native on-device down to discrete Vulkan) over
which finite impulse response (FIR) or infinite impulse response (IIR) 
and convolution-reverb filters compose with no new kernel, and
quantized matrix multiplication (\texttt{Op::DequantMatMul}). An
\texttt{Op::Custom}/\texttt{Op::CustomFn} pair provides an extension
surface so that downstream crates register new semantics without
modifying the core.

\paragraph{Three IR levels}
The operator set is expressed at three progressively lower levels that form
the compile pipeline of Figure~\ref{fig:arch}. high-level IR (HIR)
is the block vocabulary model builders emit – units such as \texttt{Linear}
and \texttt{SwiGLU} – so a transformer layer is a few nodes, not dozens of
primitives. HIR lowers to mid-level IR (MIR), the fused primitive
tensor directed acyclic graph (DAG) that is the optimizer's input: every fusion, autodiff, and
legalization pass reads and writes MIR, as does the common-IR dispatch path
of Section~\ref{sec:dispatch}. Memory planning then annotates MIR with an
arena buffer assignment to produce low-level IR (LIR), the input each backend
lowers to device code. Authoring can begin at either level – block builders
start at HIR; the lazy \texttt{Tensor} DSL traces directly to MIR – and
keeping all three as views of one operator set is what lets autodiff,
fusion, and the dispatch ladder compose without translation layers.

Two authoring surfaces target this IR: an explicit graph builder
(\texttt{Graph::new(\dots)}, \texttt{g.matmul(\dots)},
\texttt{Session::new(device).compile(g)}) and a higher-level NumPy-style
\texttt{Tensor} DSL (\texttt{rlx-tensor}) of lazy, operator-overloaded
handles – expressions such as \texttt{(\&a + \&b).relu()} trace into the
same IR are not eagerly executed, and are fused and memory-planned
only when a result is materialized.

\subsection{Optimizer}
The optimization layer comprises three crates exposed through
a single facade (\texttt{rlx-opt}).

\textit{Fusion} (\texttt{rlx-fusion}) runs region-fusion passes over the
mid-level IR and an inverse \emph{unfuse} step that re-exposes primitives
for differentiation and for backends that do not implement a fused form.
Representative fused regions include a fused attention block, fused
QK-RoPE, and a strided ``Narrow$\times3\rightarrow$Attention'' pattern
that provides the intermediate query/key/value writes.

\textit{Automatic differentiation} (\texttt{rlx-autodiff}) enables
reverse-mode \texttt{grad}, forward-mode \texttt{jvp}, Hessian-vector
products \texttt{hvp}, $n$th-order scalar derivatives, and leading-axis
batching \texttt{vmap}, all as transformations on the mid-level IR. Every
training-specific backward operator decomposes to primitives so that
derivatives can be stacked to defined order.

\textit{Compilation} (\texttt{rlx-compile}) runs legalization, memory
planning, automatic-mixed-precision policy, and post-training-quantization
insertion, producing a device-specific schedule. View-producing operators
(\texttt{Reshape}, same-dtype \texttt{Cast}, axis-0 \texttt{Narrow}) are
recognized as pure views and aliased to a root buffer with an offset, so
backends emit a no-op rather than a copy.

\paragraph{Numeric precision}
Precision is handled at the IR level.
\texttt{DType} spans floating point (F32, F16, BF16, F64, complex C64) and
integer types, with a promotion table fixing the result type of every
mixed-type operation. An \emph{automatic mixed-precision} (AMP) policy
assigns per-operator compute and accumulation types such as F32 matmul
accumulation over F16/BF16 inputs and a cast-elimination pass removes the
redundant conversions the AMP rewrite introduces. It also tracks
arithmetic intensity: F16 compute leaves a small, dispatch-bound convolutional
net unchanged but speeds a wide $16$k-hidden MLP matmul by $16\%$, accuracy
unmoved since storage and the optimizer stay F32. Below floating point,
quantization flows through the same pipeline: a post-training-quantization
(PTQ) pass inserts \texttt{FakeQuantize} nodes, quantization-aware training
(QAT) uses straight-through estimators and learned step sizes, and
weight-only quantization spans the GGUF schemes from INT8 to INT4 and
block-scaled tensor-core minifloats – the FP8/FP6/FP4 grids and arbitrary
\texttt{fNeXmY} formats, selectable per matmul (Section~\ref{sec:cases}).
Since precision is tracked in the
IR, autodiff and fusion preserve the precision, and low-precision graphs are
parity-checked against their F32 reference.

\subsection{Runtime}
The runtime (\texttt{rlx-runtime}) exposes the user-facing API:
\texttt{Session::new(device).compile(graph)} returns a
\texttt{CompiledGraph} that is executed with bound inputs. Once a graph is
pruned, placed, and scheduled, its per-device schedule is cached so that
repeated execution avoids recompilation. Multi-backend helpers
(\texttt{GraphDevices}, \texttt{DeviceRouter}, \texttt{DevicePolicy},
\texttt{FlexibleSession}) implement cost-based and environment-driven
device selection, including fallback chains across devices.

\subsection{Distributed execution}
\label{sec:distributed}
RLX scales beyond one device with two forms of parallelism over a transport
abstraction in \texttt{rlx-driver}. \emph{Pipeline parallelism} gives each
node a block of layers; only the hidden-state tensor crosses the wire,
host-side between graph runs, so no compiler change is needed and a model
too large for one device fits across several. \emph{Tensor parallelism}
shards every layer as column-sharded (gate/up) and row-sharded
(down/output) and inserts an \emph{in-graph} all-reduce
(\texttt{collective.all\_reduce}); the same all-reduce synchronizes
gradients for data-parallel \emph{training}. Collectives and point-to-point
send/receive sit on a \texttt{SymmetricTransport} trait with an in-process
\texttt{LocalTransport} and a network \texttt{NetTransport} (TCP); on Apple
Silicon the all-reduce also runs device-resident through MLX over its
\texttt{ring}/\texttt{jaccl} transports (\emph{RDMA over Thunderbolt,
fiber-optic, and PCIe}). We
evaluate for \emph{correctness}: a column-sharded and row-sharded layer reproduces the
single-node result bit-for-bit at 2- and 4-way shards, and pipeline
relaying is numerically exact. Single-box multi-process TP/PP (including
GPU-resident all-reduce) is tested; a multi-machine throughput study is
future work, as the cross-machine transport is not yet hardened.

\section{Transparent Kernel Dispatch}
\label{sec:dispatch}
A distinguishing element of RLX is how it resolves each operator to code.
Rather than silently routing unsupported operators to a CPU fallback, the
runtime classifies every operator into one of four paths: \emph{native}
(the operator is in the backend's \texttt{supported\_ops}, so a native
MSL, CUDA, CPU, \dots), and a \emph{common~IR} (a registered
logical kernel not claimed by the backend, lowered to portable primitive
MIR); \emph{rewritten} – a structural unfuse/lower yields the same semantics
in a different graph shape, and \emph{unsupported} which are still illegal after
rewriting, so the compile \emph{fails} with a diagnostic report.

The default policy, \texttt{PreferNative}, takes the native path when a
backend declares an operator in its \texttt{supported\_ops} claim set and
otherwise lowers to the common-IR path. The policy can be overridden
globally (\texttt{RLX\_KERNEL\_DISPATCH=common|native}) or per compile,
including per-operator force lists. Crucially, the \emph{same}
\texttt{supported\_ops} claim set that the system uses to decide
whether to fail a compile is also what device selection uses, so the
graph will be routed to the device only if it is guaranteed to run successfully. 
Setting \texttt{RLX\_DISPATCH\_REPORT=1} emits a
per-operator summary (native / common-IR / rewritten / missing) so that a
performance engineer can see exactly which operators are not in the fast
path, and the path to acceleration is open: implement the native
kernels, add the operator to that backend's \texttt{supported\_ops},
and re-run until the operator moves from common-IR to native.

\section{Backend Portability}
\label{sec:backends}
RLX targets fourteen runtime device variants behind one \texttt{Device}
enum, all shown in Figure~\ref{fig:arch}: \texttt{cpu}, \texttt{metal},
\texttt{mlx}, \texttt{ane}, \texttt{cuda}, \texttt{rocm}, \texttt{oneapi},
\texttt{tpu}, \texttt{hexagon}, \texttt{gpu}, \texttt{vulkan},
\texttt{opengl}, \texttt{directx}, and \texttt{webgpu}. Two more are specialty
codegen targets that consume a narrower op set and are not in the runtime
registry: \texttt{rlx-cortexm} (\texttt{no\_std} ARMv7E-M INT8) and
\texttt{rlx-fpga} (IR\,$\rightarrow$\,Verilog\,$\rightarrow$\,bitstream).

Backends share kernel sources where the hardware allows: ROCm reuses the
CUDA \texttt{.cu} sources dispatched through HIP, and the Metal and CPU
paths mirror each other's fusion patterns (e.g.\ stride-aware
Narrow$\rightarrow$RoPE). Every backend publishes a
\texttt{supported\_ops} contract stating how many of the 113+ operator kinds
it claims. For the eight production backends with published coverage tables,
CPU lowers 104+ as the reference, then the GPU backends – MLX 84, Metal 76, wgpu 75, CUDA 71, ROCm 68, ANE 57 – %
and the INT8-first TPU path claims 50 with exclusive quantized matmul/conv
operators. The \texttt{Backend} trait is open: additional runtime targets
(Cerebras wafers, mobile NPUs, Qualcomm QNN) register by implementing the
claim-set contract and providing operator thunks. Because this is the same
claim set the legalizer checks, a graph routed to a backend is guaranteed
to contain only operators that backend will compile, or the compile fails
with a report.

\section{Extensibility Case Studies}
\label{sec:cases}
Like prior unified systems~\cite{tensorflow}, RLX builds advanced features
from its own primitives rather than as core special cases.
\emph{Higher-order differentiation:} because every backward operator
decomposes to primitives, derivatives compose – RLX provides first- through
third-order scalar derivatives with inter-layer CSE, and third-order parity
holds between the CPU reference and CUDA/Metal/MLX/wgpu on $x^3$, ReLU,
tanh, GELU, and SiLU.
\emph{Quantized inference:} \texttt{Op::DequantMatMul} covers the full
\texttt{llama.cpp} scheme set (Q4\_0--Q8\_0, K-quants, IQ/TQ, MX) with GPU
dequant on Metal/CUDA/ROCm/wgpu and a fused INT4 GEMV on Metal; a
\texttt{ScaledMatMul} path adds block-scaled minifloats (FP8/FP6/FP4 and
parameterized \texttt{fNeXmY}), bit-exact against the F32 oracle on
Apple~Metal and an NVIDIA~RTX through both CUDA and native Vulkan.
\emph{Microcontroller INT8:} \texttt{rlx-cortexm} emits \texttt{no\_std}
ARMv7E-M kernels (with a companion native trainer for the INT8 weights),
outside the runtime \texttt{Backend} model and validated against reference
implementations.
\emph{Hardware synthesis:} \texttt{rlx-fpga} lowers the IR to Verilog and a
bitstream with integer-only Q0.31 requant – the same IR that drives a
server GPU also drives custom logic.

\section{Implementation and Formats}
\label{sec:impl}
\label{sec:formats}
RLX is a Cargo workspace of layered crates: the \texttt{rlx-ir} has no
backend dependencies, the optimizer crates depend only on the IR, each
backend depends on the driver and IR, and a prelude crate (\texttt{rlx})
re-exports the common surface. Backends are feature-gated and mutually
compatible, so one build can carry several and select at run time; Python
bindings (\texttt{pyrlx}, via PyO3) expose the same surface. 
RLX ingests four open-model formats, all lowering to the same IR:
\emph{safetensors} (the native path for full/mixed-precision checkpoints),
\emph{GGUF} v1--v3 (the standalone \texttt{rlx-gguf} crate covers every
\texttt{llama.cpp} scheme – Q4\_0--Q8\_0, K-quants, IQ/TQ, MX/NVFP4 – and
can also write and convert GGUF), \emph{ONNX} (imported and lowered by
\texttt{rlx-onnx} with a conformance suite, so a model targeting ONNX
Runtime compiles unchanged to any RLX backend), and \emph{rten}~\cite{rten}
(\texttt{.rten} flatbuffer models produced by the rten toolchain, sharing
import infrastructure with the ONNX path). A checkpoint from any format
is fused, quantized, memory-planned, and dispatched by the identical
pipeline.

\section{Evaluation}
\label{sec:eval}
We evaluate RLX along several axes: inference latency across RLX's backends
and other frameworks, the breadth of differentiation it can compile, model
coverage, and end-to-end correctness; we also explain what in the architecture
lets the same IR train, not just infer. (Operator coverage was quantified in
Section~\ref{sec:backends}).

\subsection{Setup}
All numbers use latest RLX on a single Apple~Silicon host. Latency is the
median (p50) of repeated trials of the encoder forward pass; the embedding
runners (RLX via \texttt{rlx-embed}, PyTorch via \texttt{transformers}, and
ONNX~Runtime over the model's ONNX export) all take identical synthetic
seq-128 inputs and report p50 on the CPU. The differentiation benchmark
uses the cycle counter (\texttt{CNTVCT\_EL0}) and is gated on thermal state
(the harness refuses to run under throttling, a real source of silent
slowdowns). Model weights come from Hugging~Face. All benchmark scripts and
result CSVs behind Fig.~\ref{fig:speedup} and the differentiation numbers
accompany the paper.

\subsection{Inference: RLX backends vs.\ other frameworks}
\begin{figure}[t]
\centering
\begin{tikzpicture}
\begin{axis}[
  width=\linewidth, height=5.6cm,
  xlabel={batch size}, ylabel={p50 latency (ms)},
  xtick={1,4,8,16,32}, xmin=0.2, xmax=33, ymin=0, ymax=135,
  grid=major, grid style={black!12},
  tick label style={font=\scriptsize}, label style={font=\scriptsize},
  legend style={font=\scriptsize, at={(0.02,0.98)}, anchor=north west,
                draw=black!30, fill=white, fill opacity=0.85, row sep=-2pt},
  legend columns=2, mark size=1.6pt, line width=0.9pt,
]
\addplot[color=mitRed,mark=*]                            coordinates {(1,1.29)(4,2.82)(8,4.83)(16,8.60)(32,16.60)};
\addplot[color=mitBrightRed,mark=square*]                coordinates {(1,8.26)(4,9.53)(8,9.74)(16,15.57)(32,28.37)};
\addplot[color=mitSilverGray!85!mitBlack,mark=triangle*] coordinates {(1,4.20)(4,8.76)(8,15.30)(16,28.44)(32,55.23)};
\addplot[color=mitBlack,mark=diamond*]                   coordinates {(1,10.08)(4,12.63)(8,22.66)(16,37.54)(32,67.48)};
\addplot[color=mitSilverGray,mark=x,mark size=2.4pt]    coordinates {(1,13.97)(4,32.15)(8,56.66)(16,112.74)(32,228.66)};
\addplot[color=blue!75!black,mark=pentagon*,dashed]  coordinates {(1,8.02)(4,13.64)(8,20.68)(16,35.80)(32,67.86)};
\addplot[color=orange!85!black,mark=pentagon,dashed] coordinates {(1,2.78)(4,4.05)(8,6.94)(16,13.75)(32,26.74)};
\addplot[color=green!55!black,mark=oplus*,dashed]    coordinates {(1,3.30)(4,14.53)(8,28.97)(16,55.19)(32,118.56)};
\addplot[color=violet!80!black,mark=oplus,dashed]    coordinates {(1,21.48)(4,62.20)(8,119.16)(16,230.33)(32,454.82)};
\legend{RLX-Metal, RLX-MLX, RLX-CPU, RLX-wgpu, RLX-Vulkan, PyTorch-CPU, PyTorch-MPS, ONNX-CPU, ONNX-CoreML}
\end{axis}
\end{tikzpicture}
\caption{One model, every engine: p50 latency (lower is better) of the five RLX
backends (solid) against PyTorch and ONNX~Runtime – each on CPU \emph{and} the
Apple GPU/ANE (dashed) – on \texttt{all-MiniLM-L6-v2}, seq-128 inputs across
batch, one Apple~Silicon host. The same compiled RLX graph runs unchanged on
every backend; RLX-Metal leads at all batches, with PyTorch-MPS the fastest
external. RLX-Vulkan (MoltenVK) and ONNX-CoreML (whose provider partitions the
graph 55 ways) run off the top above batch~8. Scripts and CSVs accompany the paper at
\url{https://github.com/MIT-RLX/rlx-paper}.}
\label{fig:speedup}
\end{figure}
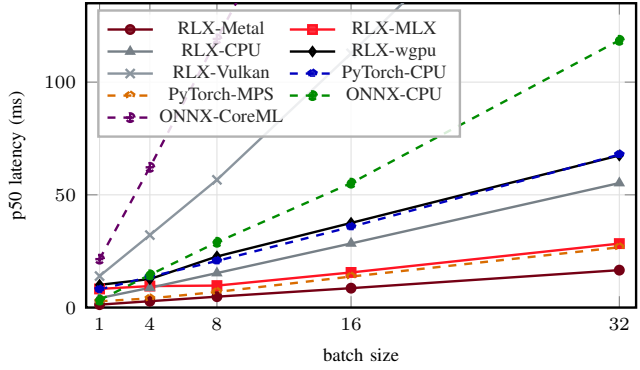

We run a \emph{single common model} – \texttt{all-MiniLM-L6-v2} (384-d, 6
layers) – through every engine identically on one Apple~Silicon host (seq~128,
batch 1-32, synthetic inputs, p50 of the encoder forward). The \emph{same}
compiled RLX graph runs unchanged on each RLX backend (CPU, Metal, MLX, and
GPU paths via wgpu/Vulkan on this host; CUDA/ROCm/oneAPI/TPU/Hexagon/ANE/
DirectX/WebGPU/OpenGL on matching hardware); PyTorch
(\texttt{transformers}) and ONNX~Runtime each run on CPU \emph{and} the Apple
GPU/ANE (MPS, CoreML). Figure~\ref{fig:speedup} plots all of them. RLX-Metal – the
router's pick – is fastest at every batch ($16.6$\,ms at batch~32, $\sim$3.3$\times$
over its own CPU path at $55.2$\,ms), ahead of the best external, PyTorch-MPS ($26.7$\,ms);
RLX's CPU path matches or beats PyTorch-CPU ($67.9$\,ms at batch~32) and leads ONNX~Runtime
($118.6$\,ms at batch~32) from batch~4, while
ONNX-CoreML trails. Scripts also wire candle as a parity reference and the
llama.cpp embedding path.

\subsection{Training: what the architecture enables}
\begin{figure}[t]
\centering
\begin{tikzpicture}
\begin{axis}[
  width=\linewidth, height=5.2cm, clip=false,
  xbar, bar width=4pt, y dir=reverse, enlarge y limits=0.05,
  xmin=0, xmax=74000, xtick={0,20000,40000,60000}, xticklabels={0,20k,40k,60k},
  scaled x ticks=false,
  xlabel={MNIST training throughput (images/s); label = wall-clock train time},
  symbolic y coords={RLX-Metal,RLX-Vk MLP,RLX g-fused,MLX-Py,TensorFlow,JAX,RLX-CUDA,RLX-MLX,burn,PyTorch-MPS,candle,RLX scalar,tinygrad},
  ytick={RLX-Metal,RLX-Vk MLP,RLX g-fused,MLX-Py,TensorFlow,JAX,RLX-CUDA,RLX-MLX,burn,PyTorch-MPS,candle,RLX scalar,tinygrad},
  tick label style={font=\scriptsize}, label style={font=\scriptsize},
]
\addplot[fill=accBk!45, draw=accBk!80, bar shift=0pt] coordinates {
  (53528,MLX-Py) (45355,TensorFlow) (41397,JAX) (19612,burn)
  (19365,PyTorch-MPS) (6930,candle) (3333,tinygrad)};
\addplot[fill=accRt, draw=accRt, bar shift=0pt] coordinates {
  (67476,RLX-Metal) (56507,RLX g-fused) (24515,RLX-MLX) (4201,RLX scalar)};
\addplot[fill=accRt, draw=accPf, line width=0.8pt, bar shift=0pt] coordinates {
  (25231,RLX-CUDA)};
\addplot[fill=accRt!35, draw=accRt, line width=0.6pt, bar shift=0pt] coordinates {
  (63967,RLX-Vk MLP)};
\node[anchor=west,font=\tiny] at (axis cs:63967,RLX-Vk MLP)   {1.9\,s};
\node[anchor=west,font=\tiny] at (axis cs:67476,RLX-Metal)    {1.8\,s};
\node[anchor=west,font=\tiny] at (axis cs:53528,MLX-Py)       {2.2\,s};
\node[anchor=west,font=\tiny] at (axis cs:56507,RLX g-fused)  {2.1\,s};
\node[anchor=west,font=\tiny] at (axis cs:45355,TensorFlow)   {2.6\,s};
\node[anchor=west,font=\tiny] at (axis cs:41397,JAX)          {2.9\,s};
\node[anchor=west,font=\tiny] at (axis cs:24515,RLX-MLX)      {4.9\,s};
\node[anchor=west,font=\tiny] at (axis cs:19612,burn)         {6.1\,s};
\node[anchor=west,font=\tiny] at (axis cs:19365,PyTorch-MPS)  {6.2\,s};
\node[anchor=west,font=\tiny] at (axis cs:25231,RLX-CUDA)     {4.8\,s};
\node[anchor=west,font=\tiny] at (axis cs:6930,candle)        {17\,s};
\node[anchor=west,font=\tiny] at (axis cs:4201,RLX scalar)    {28\,s};
\node[anchor=west,font=\tiny] at (axis cs:3333,tinygrad)      {36\,s};
\end{axis}
\end{tikzpicture}
\caption{MNIST training throughput on an identical LeNet CNN (SGD, batch~128,
2~epochs); RLX in purple, each bar labelled with its time-to-finish. Every bar
is an Apple~M4~Pro except green-edged \texttt{RLX-CUDA} (NVIDIA RTX) – the
\emph{same} IR, no rewrite, once its conv and max-pool backward run on GPU. All
reach 95--98\% accuracy; RLX trains fastest on Apple~Metal (1.8\,s), and
the top-throughput order is RLX-Metal (67{,}476\,img/s), RLX-Vk MLP
(63{,}967\,img/s), and RLX g-fused (56{,}507\,img/s). Graph-fusing
the convolution and SGD step lifts RLX's CPU path into the XLA/MLX range
($\sim$13.5$\times$ its scalar path), while RLX-Metal reaches $\sim$16$\times$
its own scalar path. The lighter \texttt{RLX-Vk MLP}
bar is the same MLP on native Vulkan (MoltenVK), the only non-CNN entry. The
chart shows portability and parity, not a claim that RLX beats a vendor-tuned
cuDNN/cuBLAS pipeline; the full training CSV (not all plotted) also has RLX
graph-fused MLP as the global top entry at 946{,}487\,img/s. Trained models
then serve inference on every RLX backend at identical 97.2\% accuracy.}
\label{fig:mnist}
\end{figure}
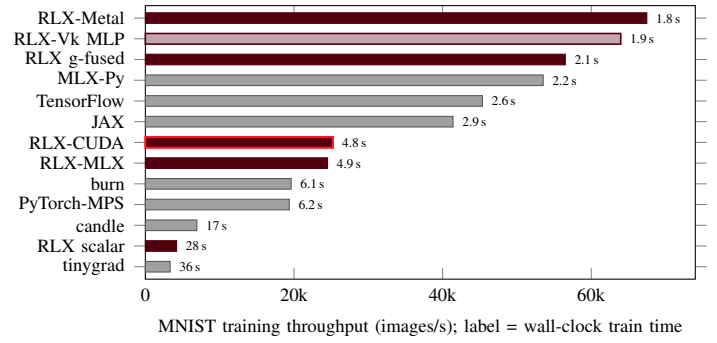

RLX trains, not only infers, and without a separate engine: gradients come
from transforming the \emph{same} IR. Three choices carry this.
\texttt{rlx-autodiff} rewrites the HIR graph in place, so reverse-, forward-,
and higher-order gradients (\S\ref{sec:hoad}) compile through the same fusion,
legalization, and device-routing pipeline as the forward pass; the optimizer
is part of the graph (an SGD/Adam step fuses into the schedule with weights resident via JIT);
and the fusion levers that speed inference (chain fusion, an im2col${+}$BLAS convolution) 
minimize the backward pass identically. The payoff is empirical (Fig.~\ref{fig:mnist}): on
an identical LeNet CNN, graph-fusing the convolution and the SGD step lifts
RLX's CPU training to the throughput of the XLA/MLX stacks. In
Fig.~\ref{fig:mnist}, the top-throughput order is RLX-Metal
(67{,}476\,img/s), RLX-Vk MLP (63{,}967\,img/s), and RLX g-fused
(56{,}507\,img/s), directly from the plotted values in Fig.~\ref{fig:mnist}.
The fastest RLX path is $\sim$16$\times$ its own
scalar path, while
every framework lands in the same 95--98\% accuracy
band. The fused path carries no overhead even versus hand-written code: an MLP
reaches 946{,}487\,img/s (top row in the full table), ahead of
NumPy${+}$BLAS (787{,}349\,img/s) at higher accuracy.
Because backward is just more IR, the trained model then serves
inference on \emph{every} RLX backend at identical accuracy – up to
161{,}660\,img/s on the Neural Engine – and emits the INT8 microcontroller
deployment of \S\ref{sec:cases}. On-device ANE training stays efficient by
binding trainable weights as graph \emph{inputs} (compile once, predict per
step). The gradients match the CPU bit-for-bit, CPU and Metal
are the most exercised, and some larger convolutional backward graphs remain
bounded by Core ML lowering.

\subsection{Higher-order differentiation}
\label{sec:hoad}
To exercise the autodiff transforms beyond inference, we benchmark
third-order scalar differentiation of a cubic-sum objective (batch~200) on
the Apple~Silicon host (median of 50 runs, via the accompanying \texttt{bench/}
harness). At these sizes the CPU is fastest (median time grows $\sim O(N^2)$,
$0.5$\,µs at $N{=}64$ to $12.7$\,µs at $N{=}4096$) while the launch-bound Metal
and MLX paths stay nearly flat, overtaking only at much larger $N$. The point
is that the \emph{same} composable transforms differentiate to third order
across CPU, Metal, and MLX (and, elsewhere, wgpu and CUDA) at on-device parity
with the CPU.

\subsection{Supported models}
\label{sec:zoo}
A companion repository \url{https://github.com/MIT-RLX/rlx-models} provides roughly 100+ model-family crates – LLMs
(Qwen3, LLaMA~3.2, Gemma, Phi, GLM), state-space models (Mamba, LFM),
vision and diffusion (SAM, DINOv2, BioCLIP-2, FLUX.2), embeddings (BERT,
NomicBERT), and automatic speech recognition (ASR) and text-to-speech (TTS) models 
(Whisper, TinyTTS, Orpheus, Voxtral) – all at reference parity.
Qwen3 uses safetensors \emph{and} GGUF at 100\% top-1 parity with
Hugging~Face; BioCLIP-2 reaches 100\% parity on CPU/Metal/MLX/wgpu; the
Cortex-M INT8 path reaches 96.6\% on nRF52840; and the TPU path matches
MiniLM-L6 through PJRT – one IR from a microcontroller to a server
accelerator.

\begin{figure}[t]
\centering
\begin{tikzpicture}[font=\tiny]
\begin{axis}[
  width=\linewidth, height=5.7cm,
  xlabel={deployment reach (platforms, of 12)},
  ylabel={capability depth (of 4)},
  xmin=-0.3, xmax=13, ymin=-0.25, ymax=5,
  xtick={0,2,4,6,8,10,12}, ytick={0,1,2,3,4},
  grid=both, grid style={black!8},
  tick label style={font=\scriptsize}, label style={font=\scriptsize},
  clip=false,
]
\addplot[only marks, mark=*, mark size=1.2pt, color=accBk] coordinates {
 (4,1.5)(6.5,2.5)(2,2)(3,1.55)(5,0.5)(6,1.5)(4.4,2)(6,3)(3,1)(4.6,1.4)
 (4.3,0.45)(4.7,0.6)(3.1,1.15)(5.5,1.6)(2.4,2.12)(2.75,1.32)(2.5,0.9)(4.85,2.05)(2.95,1.92)};
\node[anchor=east] at (axis cs:4,1.5) {PyTorch};
\node[anchor=east] at (axis cs:6.5,2.5) {TensorFlow};
\node[anchor=east] at (axis cs:2,2) {JAX};
\node[anchor=south east] at (axis cs:3,1.6) {MLX};
\node[anchor=north] at (axis cs:5,0.5) {Core ML};
\node[anchor=west] at (axis cs:6,1.5) {ONNX RT};
\node[anchor=east] at (axis cs:4.4,2) {candle};
\node[anchor=east] at (axis cs:6,3) {burn};
\node[anchor=east] at (axis cs:3,1) {tch};
\node[anchor=west] at (axis cs:4.6,1.4) {rten};
\node[anchor=east] at (axis cs:4.3,0.45) {ggml};
\node[anchor=west] at (axis cs:4.7,0.6) {llama.cpp};
\node[anchor=west] at (axis cs:3.1,1.15) {mistral.rs};
\node[anchor=south] at (axis cs:5.5,1.6) {IREE};
\node[anchor=south] at (axis cs:2.4,2.12) {Glow};
\node[anchor=west] at (axis cs:2.75,1.32) {MAX};
\node[anchor=west] at (axis cs:2.5,0.9) {TensorRT};
\node[anchor=west] at (axis cs:4.85,2.05) {tinygrad};
\node[anchor=south west] at (axis cs:2.95,1.92) {Luminal};
\node[anchor=center, color={red}] at (axis cs:12,4) {RLX};
\end{axis}
\end{tikzpicture}
\caption{Deployment reach (twelve platforms) versus capability depth (four: native-Rust author{+}compile, fused compiler IR, training-grade autodiff, and MCU/FPGA codegen). RLX occupies the top-right corner; CoreML and MLX cover the Apple family but stop there.}
\label{fig:compare}
\end{figure}
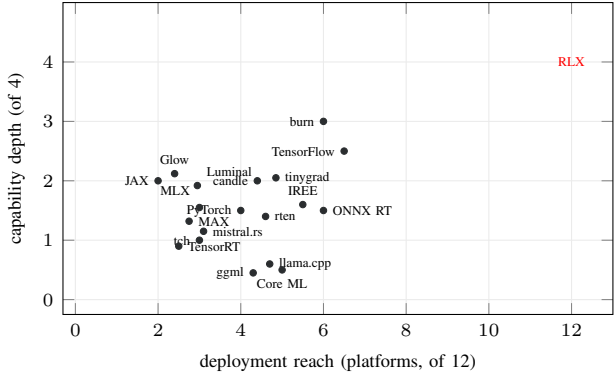

\section{Comparison with ML Frameworks,\\Apple Libraries, and Inference Engines}
\label{sec:comparison}
We place RLX within the field (Fig.~\ref{fig:compare}): Python frameworks
(PyTorch~\cite{pytorch}, TensorFlow~\cite{tensorflow}, JAX~\cite{jax});
eager kernel libraries (candle~\cite{candle}, tch~\cite{tch},
MLX~\cite{mlx}); inference engines (rten~\cite{rten},
ONNX~Runtime~\cite{ort,onnx}, CoreML~\cite{coreml}); LLM stacks
(ggml~\cite{ggml}, llama.cpp~\cite{llamacpp}, mistral.rs~\cite{mistralrs});
burn~\cite{burn}; and compiler stacks (IREE~\cite{iree}, Glow~\cite{glow},
MAX~\cite{mojo}, TensorRT~\cite{tensorrt}, tinygrad~\cite{tinygrad},
Luminal/ZML~\cite{luminal,zml}).

Each covers part of the design space but does not implement it end-to-end as RLX does.
The majority of libraries and LLM stacks carry no whole-program compiler
IR with cross-backend lowering; the inference engines and most compiler
stacks are inference-only, without first-class differentiation; the Apple
libraries are Apple-only; tch and TensorRT bind a heavy or single-vendor
runtime; and all but candle, burn, rten, mistral.rs, and Luminal are
non-Rust. Among the compilers only JAX and TensorFlow match RLX's
transform-based, fused-IR view of differentiation – but neither is native
Rust or have lower overhead or better performance, 
nor reaches the Neural Engine, microcontrollers, or FPGAs from one IR.

RLX alone combines a native-Rust author-and-compile path, a primitive-level
IR with reverse/forward/higher-order differentiation plus batching,
transparent multi-backend dispatch, and edge codegen with full platform
reach. MLX and CoreML are themselves RLX backends, so a model authored once
targets Apple Silicon \emph{and} NVIDIA, AMD, TPU, web, and MCU – no
rewrite. Tensor-/pipeline-parallel distributed execution
(Section~\ref{sec:distributed}) extends the advantage beyond what the Rust
neighbors provide.

\section{Conclusion and Future Work}
\label{sec:conclusion}
We presented RLX, a tensor compiler and runtime that unifies the
graph-compiler and kernel-runtime roles in one Rust workspace. A
primitive-level, three-level IR feeds a fusion, autodiff, and
precision optimizer; a runtime resolves every operator through a
transparent dispatch ladder and routes it across fourteen runtime targets
plus microcontroller and FPGA codegen, with parallel
distributed execution over TCP and RDMA. Across a comparison spanning the
Python frameworks, Apple libraries, inference engines, LLM stacks, and
retargetable compilers, RLX is distinguished by uniting a primitive-level
compiler IR with training-grade differentiation, transparent dispatch, and
edge codegen in one native-Rust tree – and, in the reported benchmarks,
posts the top throughput results (fastest inference
from batch~4 up on a common embedding model, and the top training-table entry
at 946k\,img/s). Future work
will include GPU and TPU optimizations, better training, and better
multi-machine transport.

\section*{Acknowledgments}
The authors thank the maintainers of the open-source Rust and ML ecosystems on
which RLX builds, and the reviewers for their comments. Kimi-K3 LoRA pretrained 
on \textbf{rlx} and \textbf{rlx-models} codebases  running on \textbf{rlx} inference 
engine was used to produce an initial draft  of the paper that was subsequently 
significantly revised by human authors.

\end{document}